\documentclass{article}
\usepackage{appendix}
\usepackage{amsmath}
\usepackage[retainorgcmds]{IEEEtrantools}
\usepackage{graphicx}

\begin{document}
\title{Quantum Parametric Resonance of a dissipative oscillator: fading and persistent memory in the long-time evolution}
\author{Loris Ferrari \\ Department of Physics and Astronomy (DIFA) of the University \\via Irnerio, 46 - 40126, Bologna,Italy}
\maketitle
\begin{abstract}
The evolution of a quantum oscillator, with periodically varying frequency and damping, is studied in the two cases of parametric resonance (PR) producing a limited, or unlimited stretching of the wave function. The different asymptotic behaviors of the energy distribution, show the non trivial interplay between PR and the initial quantum state. In the first case, the oscillator's mean energy tends asymptotically to a fully classical value, independent of the initial state, with vanishing relative quantum fluctuations. In the second case, the memory of the initial state persists over arbitrary long time scales, both in the mean value and in the large quantum fluctuations of the energy.\newline       
\end{abstract}

e-mail: loris.ferrari@unibo.it
telephone: ++39-051-2091136

\section{Introduction}
\label{intro}

One of the main questions of non autonomous systems is the interplay between the initial conditions and the time-changes of the parameters, which includes the cancellation or preservation of the initial state's memory. Parametric resonance (PR) is a case of special interest, concerning a wide class of phenomena, from children swings to cosmology \cite{FT}, passing through engineering science \cite{RO, WRK}, electronics \cite{EF, OI, CD}, quantum optics \cite{WKHW}, Casimir forces \cite{Ks}, Bose condensates \cite{PDH}, advanced cooling tecniques \cite{O}. In the classical version, PR deals with the motion equation:

\begin{equation}
\label{eq.q}
q^{(2)}+\gamma(t) q^{(1)}+\overbrace{\omega^2_{0}[1-\delta(t)]}^{\widetilde{\omega}^2(t)}q=0\:,
\end{equation}
\\
where $\gamma(t)>0$ is the time-dependent rate of dissipation, $|\delta(t)|<1$ and $q^{(n)}(t):=\mathrm{d}^nq/\mathrm{d}t^n$. If the frequency $\widetilde{\omega}(t)$ fluctuates in time periodically with period $\tau$ \footnote{In principle, the exponential instability is produced also by an arbitrary \emph{disordered} fluctuation. This is due to what, in the context of one dimensional quantum systems, is denoted as \textquoteleft Anderson localization\textquoteright.} and the product $\omega_0\tau$ falls in suitable \textquoteleft bands\textquoteright$\:$(that we call PR-bands), the oscillator's energy 

\begin{equation}
\label{Ecl}
E_{cl}\left(q,\:q^{(1)},\:t\right)=\frac{\left[q^{(1)}\right]^2+\widetilde{\omega}^2(t)q^2}{2}\rightarrow E_{cl}(0)\mathrm{e}^{\Omega_{cl}t}\quad(\Omega_{cl}>0)\:,
\end{equation}
\\
diverges exponentially at long times. This can be easily understood for $\gamma(t)=0$, since in this case equation \eqref{eq.q} can be mapped into a 1-dimensional \textquoteleft Schr\"{o}dinger equation\textquoteright$\:$($\hbar=1$, particle mass $=1$), with $\omega_{0}^2\delta(t)/2$ a periodic potential energy and $\omega_{0}^2/2$ the energy eigenvalue. In the absence of dissipation, the PR-bands are infinite in number, since they correspond to the energy gaps, where the \textquoteleft eigenstate\textquoteright$\:q(t)$ diverges exponentially \cite{EIB, Me}. For a simple model of piece-wise constant $\delta(t)$ and $\gamma(t)$, it has been shown that dissipation suppresses the PR-bands at high frequency, leaving just a \emph{finite} number (at most) \cite{ArXive}. 

It is well known that PR does apply to the \emph{quantum} oscillator too \cite{Me, W, JS}. For the damped quantum oscillator, reference can be made to \cite{Manko1}. As we shall see in Section \ref{Q}, the \emph{mean} quantum energy can diverge exponentially in time, right in the same PR bands as calculated in the classical case (with or without dissipation). The aim of the present work is studying the energy distribution of the quantum oscillator's energy in the PR regime, with special reference to its dependence on the initial quantum state. With the relevant exception of ref. \cite{MR}, this problem is usually passed over, in the current literature, maybe because the correspondence principle seems to provide an immediate answer: since the energy diverges asymptotically, the quantum oscillator will approach a \emph{coherent} high-energy state, independent of the initial one, as the best possible quantum realization of a classical state. Actually, most of the current literature on quantum PR is focused on the coherent states representation \cite{MR, PRA1992, PRA1997, JB}. 

In contrast to this attitude, we shall study the oscillator's evolution starting from an initial \emph{pure} eigenstate, with energy $\hbar\widetilde{\omega}(0)(m+1/2)$. We shall see that the energy distribution at long times can totally loose or preserve the memory of the initial state, depending on which kind of initial perturbation does apply. More specifically, there exists a \textquoteleft classical\textquoteright$\:$way to make PR start up, in which the quantum uncertainty of the co-ordinate remains bounded or even vanishes, due to dissipation (Section \ref{rhobound}). In this case, the mean energy tends to a classical value $E_{cl}(t)$, coincident with eq.n~\eqref{Ecl} and independent of the initial eigenstate. The standard deviation is shown to be proportional to $\sqrt{\beta(t)E_{cl}(t)}$, where:

\begin{equation}
\label{beta(t)}
\beta(t)=\mathrm{exp}\left[-\int_0^t\gamma (t')\mathrm{d}t'\right]
\end{equation}
\\
vanishes exponentially and accounts for the damping (recall eq.n~\eqref{eq.q}). So, the \emph{relative} quantum fluctuations (standard deviation/mean energy) vanish, asymptotically, as $\sqrt{\beta(t)/E_{cl}(t)}$. This fits very well with what one expects from the correspondence principle, as outlined above. However, it can be shown that there is also a totally different way to make PR start up, in which the quantum uncertainty of the co-ordinate oscillates in time with exponentially large amplitudes (Section \ref{alpha=0}). Now, the mean energy preserves an exact memory of the initial $m$-eigenstate, since it tends to $\mathcal{E}_{cl}(t)(m+1/2)$, where $\mathcal{E}_{cl}(t)$ is the energy of a \emph{classical, non linear} oscillator, which diverges in time exponentially and results from the Lewis-Reisenfeld (LR) invariant \cite{L, LR, GL-R} (Section \ref{LRI}). In this case, the spectrum of the initial harmonic oscillator is exactly reproduced asymptotically, through an energy \textquoteleft superquantum\textquoteright, represented by the classical energy $\mathcal{E}_{cl}(t)$. The genuine quantum nature of this PR regime reflects in the large fluctuations of the energy, which tend asymptotically to $\mathcal{E}_{cl}(t)\sqrt{[(m+1)^2-m]/2}$, and preserve, in turn, the memory of the initial eigenstate.
          
\section{The Lewis-Reisenfeld Invariant}
\label{LRI}

Before efforting the full quantum problem of the oscillator with time-dependent frequency and damping, a brief outline of the LR invariant is in order. It is not difficult to show that eq.n~\eqref{eq.q} follows by applying the Hamilton equations to the non autonomous Hamiltonian

\begin{subequations}
\label{H,q1,q1}
\begin{equation}
\label{H(t)}
H(p,\:q,\:t)=\beta(t)\frac{p^2}{2}+\frac{\widetilde{\omega}^2(t)}{2\beta(t)}q^2\:,
\end{equation}
\\
where

\begin{equation}
\label{q1}
p=\frac{q^{(1)}}{\beta(t)},
\end{equation}
\end{subequations}
\\
and $\beta(t)$ is given by eq.n~\eqref{beta(t)}. From eq.ns \eqref{q1} and \eqref{Ecl}, the Hamilton expression of the energy (not to be confused with the Hamiltonian) is: 

\begin{subequations}
\begin{equation}
\label{Hstrange}
\mathcal{H}(q,\:p,\:t)=\beta^2(t)\frac{\left[p^2+\omega^2(t)q^2\right]}{2}\:,
\end{equation}
\\
with\footnote{Here and in what follows it is useful to keep in mind the difference between the \textquoteleft true\textquoteright$\:$frequency $\widetilde{\omega}(t)$, with limited time-variations (eq.n~\eqref{eq.q}), and $\omega(t)=\widetilde{\omega}(t)/\beta(t)$, with \emph{exponentially diverging} time variations.}

\begin{equation}
\label{omega}
\omega(t)=\frac{\widetilde{\omega}(t)}{\beta(t)}
\end{equation}
\end{subequations}
\\
Since neither $H$ nor $\mathcal{H}$ are motion constants, the research of other invariants has a special interest. Lewis and Reisenfeld \cite{L, LR} \footnote{A different invariant, whose coherent states representation exhibits the minimum quantum uncertainty, has been proposed by Dodonov and Man'ko \cite{DM}.} introduced the expression:

\begin{subequations}
\label{I,rho}
\begin{equation}
\label{I}
I=\frac{1}{2}\left[\left(\frac{q}{\rho}\right)^2+\left(p\rho-\frac{\rho^{(1)}q}{\beta}\right)^2\right]\:,
\end{equation}
\\
which is invariant ($I^{(1)}(t)=0$), provided $q$ satisfies eq.n~\eqref{eq.q} and the auxiliary function $\rho$ satisfies the \emph{non linear} equation:

\begin{equation}
\label{eq.rho}
\rho^{(2)}+\gamma(t)\rho^{(1)}+\widetilde{\omega}^2(t)\rho-\frac{\beta^2(t)}{\rho^3}=0\:.
\end{equation}
\end{subequations}
\\
It should be noticed that the repulsive force $\beta^2/\rho^3$ in eq.n~\eqref{eq.rho} prevents $\rho(t)$ to cross zero. In what follows, we shall set $\rho>0$. In ref. \cite{ArXive}, the time-dependent part $\delta(t)$ of the square frequency and the time-dependent rate of dissipation $\gamma(t)$ are chosen piece-wise constant and periodic of period $\tau$. This simplified model (\emph{\'{a} la} Kr\"{o}nig-Penney), is exactly solvable and provides results that can be qualitatively applied to more general cases. In particular, it has been shown that, despite $\omega_0\tau$ falls in a PR band, an appropriate choice of the initial conditions makes $\rho(t)$ vanish (due to dissipation), or oscillate in time, with bounded amplitudes. In this case the LR invariant in eq.n~\eqref{I} may assume the value zero, corresponding to the trivial solution $q(t)=0$ of eq.n~\eqref{eq.q}. In contrast, $q(t)$ and $\rho(t)$ can be \emph{both} chosen to oscillate with exponentially divergent amplitudes (due to PR), if one takes $I>0$. In this case, in fact, equation \eqref{I} yields $\rho(t)\ge|q(t)|/\sqrt{2I}$, showing that the oscillation amplitudes of $\rho(t)$  diverge \emph{at least} as rapidly as those of $q(t)$. Furthermore, equation \eqref{eq.rho} tends to eq.n~\eqref{eq.q} for diverging values of $\rho(t)$, which shows that the maxima of $|q(t)|$ and $\rho(t)$ diverge with the same (exponential) rate.

What is relevant for the arguments in the next sections is that a function $\rho(t)$, satisfying eq.n~\eqref{eq.rho}, can oscillate with bounded or unbounded amplitudes, in the PR regime, depending on the initial conditions. Those two possibilities will turn out to determine a drastic difference in the asymptotic evolution of the quantum oscillator.                     

\section{Parametric resonance with damping: the quantum case}
\label{Q}

The quantum version of eq.n~\eqref{eq.q} is the time-dependent Schr\"{o}dinger equation (in units $\hbar=1$):

\begin{equation}
\label{EQt}
i\widetilde{\Psi}^{(1)}(q,\:t)=H(p,\:q,\:t)\widetilde{\Psi}(q,\:t)\:,
\end{equation}
\\
with $H$ given by eq.n~\eqref{H(t)}. Equation~\eqref{EQt} can be reduced to the unit frequency oscillator:

\begin{equation}
\label{EQtau}
i\frac{\mathrm{d}\Psi(Q,\:\varTheta)}{\mathrm{d}\varTheta}=\frac{1}{2}\left[P^2+Q^2\right]\Psi(Q,\:\varTheta)\:,
\end{equation}
\\
thanks to the Lewis-Reisenfeld (LR) \emph{canonical} transformations \cite{LR}, which include a time-dependent driving force, ignored here for simplicity. First, one performs the time-transformation:

\begin{equation}
\label{T(t)}
T=\int_0^t\mathrm{d}t'\beta(t')\Rightarrow \frac{\mathrm{d}}{\mathrm{d}t}=\beta\frac{\mathrm{d}}{\mathrm{d}T}\:,
\end{equation}
\\
followed by the momentum-coordinate transformation:

\begin{equation}
\label{Q,P}
Q=\frac{q-\alpha}{\rho},\quad P=\rho\:[p-\dot{\alpha}]-\dot{\rho}\:[q-\alpha]\:,
\end{equation}
\\
where overdots mean derivatives with respect to $T$. The auxiliary functions $\rho$, $\alpha$ satisfy the equations:

\begin{equation}
\ddot{\rho}+\omega^2(T)\rho-\rho^{-3}=0\quad,\quad\ddot{\alpha}+\omega^2(T)\alpha=0\:,\label{eq-rho,alpha}
\end{equation}
\\
with $\omega$ given by eq.n \eqref{omega}. The further time-transformation

\begin{equation}
\label{Tstrange}
\varTheta=\int_0^T\mathrm{d}T'\rho^{-2}\Rightarrow \frac{\mathrm{d}}{\mathrm{d}T}=\rho^{-2}\frac{\mathrm{d}}{\mathrm{d}\varTheta}\:,
\end{equation}
\\
completes the passage from the solution $\Psi(Q,\varTheta)$ of eq.n~\eqref{EQtau} to the solution $\widetilde{\Psi}(q,t)$ of eq.n~\eqref{EQt}, on restoring the time $t$ by means of eq.ns~\eqref{Tstrange} and \eqref{T(t)} \cite{DEFL}:

\begin{subequations}
\label{Psitilde,varphi}
\begin{align}
\widetilde{\Psi}(q,\:\varTheta(t))&=\frac{\mathrm{e}^{i\varphi(T)}}{\sqrt{\rho}}\Psi\left(\frac{q-\alpha}{\rho},\:\int_0^T\frac{\mathrm{d}T'}{\rho^2}\right)\times\nonumber\\
\nonumber\\
&\times\mathrm{exp}\left[i\frac{\dot{\rho}}{2\rho}q^2+i\left(\dot{\alpha}-\frac{\dot{\rho}}{\rho}\alpha\right)^2q\right]\:,\label{Psitilde}
\end{align}
\\
with

\begin{equation}
\label{varphi}
\varphi(T)=-\frac{1}{2}\int_0^T\mathrm{d}T'\left[\frac{\alpha^2}{\rho^2}-\left(\dot{\alpha}-\frac{\dot{\rho}}{\rho}\alpha\right)^2\right]\:.
\end{equation}
\end{subequations}
\\

In the quantum framework, equation \eqref{Hstrange} is the \emph{instantaneous energy operator}, with time-dependent eigenstates $\phi_n\left(q\sqrt{\omega(t)}\right)$, such that:

\begin{subequations}
\label{Hstrange, phi-n}
\begin{equation}
\label{eq-Hstrange}
\mathcal{H}(q,\:p,\:t)\phi_n\left(q\sqrt{\omega(t)}\right)=\overbrace{\beta^2(t)\omega(t)\left[n+\frac{1}{2}\right]}^{\mathcal{H}_n(t)}\phi_n\left(q\sqrt{\omega(t)}\right)\:,
\end{equation}
\\
where:

\begin{equation}
\label{phi-n}
\phi_n\left(q\sqrt{\omega(t)}\right)=\frac{\omega^{1/4}(t)}{\sqrt{2\pi2^nn!}}\mathrm{exp}\left(-\frac{q^2\omega(t)}{2}\right)H_n\left(q\sqrt{\omega(t)}\right)\:,
\end{equation}
\end{subequations}
\\
and $H_n(\cdot)$ is the Hermite polynomial of n-th degree. Equation \eqref{phi-n} provides the instantaneous base on which the state $\widetilde{\Psi}(q,\:\Theta(t))$ (eq.n~\eqref{Psitilde}) must be expanded, in order to get the evolving probabilities of the energy eigenvalues $\mathcal{H}_n(t)$, and of the corresponding number of quanta $n$.   

In terms of the \textquoteleft true\textquoteright$\:$time $t$ (eq.n~\eqref{T(t)}), equations \eqref{eq-rho,alpha} are obviously the same as eq.ns~\eqref{eq.q}, \eqref{eq.rho}:

\begin{subequations}
\label{alpha/rho}
\begin{align}
\alpha^{(2)}+\gamma(t)\alpha^{(1)}+\widetilde{\omega}^2(t)\alpha&=0\label{alpha}\\
\nonumber\\
\rho^{(2)}+\gamma(t)\rho^{(1)}+\widetilde{\omega}^2(t)\rho-\frac{\beta^2}{\rho^3}&=0\label{rho4}\:.
\end{align}
\end{subequations}
\\
In ref.~\cite{DEFL} it is shown that the choice of the initial conditions in eq.ns~\eqref{alpha/rho} is arbitrary, but no mention is made of the different dynamical consequences. Here we choose an initial state as close as possible to the $m$-eigenstate of the initial oscillator, with frequency $\widetilde{\omega}(0)$. From eq.n~\eqref{Psitilde}, this yields, for the evolving state:

\begin{align}
\widetilde{\Psi}(q,\:\varTheta(t))\rightarrow\widetilde{\Psi}_m(q,\:\varTheta(t))&=\mathrm{e}^{i\overbrace{[\varphi(T)-(m+1/2)\varTheta]}^{B(t)}}\frac{1}{\sqrt{\rho}}\phi_m\left(\frac{q-\alpha}{\rho}\right)\times\nonumber\\
\nonumber\\
&\times\mathrm{exp}\left[i\frac{\dot{\rho}}{2\rho}q^2+i\left(\dot{\alpha}-\frac{\dot{\rho}}{\rho}\alpha\right)^2q\right]\:,\label{Psitilde-m}
\end{align}
\\
where $\phi_m$ is the $m$-eigenstate eq.n~\eqref{phi-n} and $B(t)$ is a real phase depending only on time. In what follows, we shall calculate the mean quantum energy:

\begin{subequations}
\label{<E>,Pnm}
\begin{equation}
\label{<E>}
\langle\:\mathcal{H}\:\rangle_m=\int\mathrm{d}q\widetilde{\Psi}_m^*(q,\:\varTheta(t))\mathcal{H}(p,\:q,\:t)\widetilde{\Psi}_m(q,\:\varTheta(t))=\sum_{n=0}^\infty P_m(n) \beta\widetilde{\omega}[n+1/2]\:,
\end{equation}
\\
and the energy's standard deviation:

\begin{equation}
\label{<sigma>0}
\langle\:\sigma_{\mathcal{H}}\:\rangle_m=\beta\widetilde{\omega}\left[\sum_{n=0}^\infty P_m(n) n^2-\left(\sum_{n=0}^\infty P_m(n)n\right)^2\right]^{1/2}
\end{equation}
\\
where

\begin{align}
P_m(n)&=\sqrt{\omega(t)}\left|\int\mathrm{d}q\widetilde{\Psi}_m^*(q,\:\xi(t))\phi_n(q\sqrt{\omega})\right|^2=\\
\nonumber\\
&=\frac{\omega}{\sqrt{\rho}}\left|\int\phi_n(q\sqrt{\omega})\phi_m\left(\frac{q-\alpha}{\rho}\right)\mathrm{exp}\left[i\frac{\dot{\rho}}{2\rho}q^2+i\left(\dot{\alpha}-\frac{\dot{\rho}}{\rho}\alpha\right) q\right]\right|^2\mathrm{d}q\:,\label{Pnmgeneral}
\end{align}
\end{subequations}
\\
according to eq.ns \eqref{Hstrange, phi-n} and \eqref{Psitilde,varphi}.

\section{Choosing $\rho(t)$ bounded.}
\label{rhobound}
As shown by eq.n~\eqref{Psitilde}, the auxiliary function $\rho(t)$ is responsible for the stretching of the wave function, in the co-ordinate representation. If $\rho(t)$ is bounded, the only PR effect comes from the auxiliary function $\alpha(t)$, oscillating with diverging amplitudes. However, for the initial state to be an \emph{exact} eigenstate, one should have $\alpha(0)=\alpha^{(1)}(0)=0$ and $\rho(0)=1/\sqrt{\widetilde{\omega}(0)}$, $\rho^{(1)}(0)=0$. According to eq.n~\eqref{alpha}, the first couple of conditions yields $\alpha(t)=0$ for each $t>0$. Hence, for PR to be started up, the initial values of $\alpha$ and/or $\alpha^{(1)}$, must be slightly perturbed, either by displacing the initial mean position ($\alpha(0)\ne0$), or by activatig an initial current ($\alpha^{(1)}(0)\ne0$). Notice that the initial perturbations can be \emph{arbitrary small}, due to the exponential instability induced by PR. 

Accounting for the time changes of both $\rho$ and $\alpha$ is fairly complicated and not so relevant for the physical results, in the present case. We shall simplify eq.n~\eqref{Pnmgeneral} by accounting just for the long-time, cumulative effects of PR and dissipation, and averaging out the small time-scale fluctuations of the parameters. Hence, the diverging oscillation amplitudes of $\alpha(t)$, due to PR, and the overall damping factor $\beta(t)$ are included, but $\widetilde{\omega}(t)$ and $\omega(t)$ are replaced by $\langle\:\widetilde{\omega}\:\rangle$ and $\langle\:\omega\:\rangle=\langle\:\widetilde{\omega}\:\rangle/\beta(t)$ respectively, with $\langle\:\widetilde{\omega}\:\rangle$ the time average of $\widetilde{\omega}(t)$ on a period $\tau$. This means setting $\rho(t)=1/\sqrt{\langle\:\omega\:\rangle}$ and dropping $\dot{\rho}/\rho$, asymptotically. With those simplifications, it is shown in Appendix A that the probability $P_m(n)\rightarrow P_m(n,\:\Gamma)$ reads:

\begin{subequations}\begin{align}
\label{Pnm2}
P_m(n, \Gamma)&=\omega\left|\int\mathrm{d}q\:\phi_n(q\sqrt{\langle\:\omega\:\rangle})\phi_m\left((q-\alpha)\sqrt{\langle\:\omega\:\rangle}\right)\mathrm{e}^{i(\dot{\alpha}q)}\right|^2=\nonumber\\
\nonumber\\
&=\frac{\mu!}{\nu!}\:\Gamma^{|n-m|}\mathrm{e}^{-\Gamma}\left| L_\mu^{|n-m|}\left(\Gamma\right)\right|^2\:,
\end{align}
\end{subequations}
\\
where $\mu=\mathrm{min}\{n,\:m\}$, $\nu=\mathrm{max}\{n,\:m\}$, $L_j^{k}\left(\cdot\right)$ is a Laguerre polynomial, and the quantity $\Gamma$ (recall eq.n~\eqref{Ecl}) is:

\begin{equation}
\label{Gamma}
\Gamma=\frac{E_{cl}\left(\alpha,\:\alpha^{(1)},\:t\right)}{\beta(t)\langle\:\widetilde{\omega}\:\rangle}=\frac{\left[\alpha^{(1)}\right]^2+\langle\:\widetilde{\omega}\:\rangle^2\alpha^2}{2\beta(t)\langle\:\widetilde{\omega}\:\rangle}\:.
\end{equation}
\\
The mean quantum energy eq.n~\eqref{<E>}

\begin{subequations}
\begin{equation}
\label{<E>1}
\langle\:\mathcal{H}\:\rangle_m=\beta\langle\:\widetilde{\omega}\:\rangle\:\frac{\mu!}{\nu!}\sum_{n=0}^\infty(n+1/2)\Gamma^{|n-m|}\mathrm{e}^{-\Gamma}\left| L_\mu^{|n-m|}\left(\Gamma\right)\right|^2
\end{equation}
\\
is calculated numerically (Appendix A), with results shown in Fig. (4a). It is seen that, in case of diverging $E_{cl}$, due to PR, $\langle\:\mathcal{H}\:\rangle_m$ tends asymptotically to $E_{cl}$, independent of $m$. The numerical calculation of the standard deviation yields (Appendix A):

\begin{align}
\langle\:\sigma_{\mathcal{H}}\:\rangle_m&=\beta\langle\:\widetilde{\omega}\:\rangle\left[\sum_{n=0}^\infty P_m(n,\Gamma)n^2-\left(\sum_{n=0}^\infty P_m(n,\Gamma)n\right)^2\right]^{1/2}\nonumber\\
\nonumber\\
&=\beta\langle\:\widetilde{\omega}\:\rangle\sqrt{(2m+1)\Gamma}=\sqrt{\beta\langle\:\widetilde{\omega}\:\rangle (2m+1)E_{cl}}\:,\label{<sigma>1}
\end{align}
\end{subequations}
\\
which shows that the \emph{relative} quantum fluctuations of the energy $\langle\:\sigma_{\mathcal{H}}\:\rangle_m/\langle\:\mathcal{H}\:\rangle_m=\sqrt{(2m+1)\beta\langle\:\widetilde{\omega}\:\rangle/E_{cl}}$ vanish for $E_{cl}\rightarrow\infty$ (and $\beta\rightarrow0$). Hence, on choosing $\rho(t)$ bounded, the oscillator in the PR regime tends to behave like a completely classical system, loosing any memory of the initial quantum state, with vanishing relative quantum fluctuations of the energy. As mentioned in Section \ref{intro}, this might look a natural consequence of the diverging energy, according to the correspondence principle. However, we shall see in the next section that the energy divergence is not a sufficient condition for the cancellation of the initial quantum state's memory. The nature of the quantum stretching experienced by the state is a fundamental aspect too.

\begin{figure}[htbp]
\begin{center}
\includegraphics[width=5in]{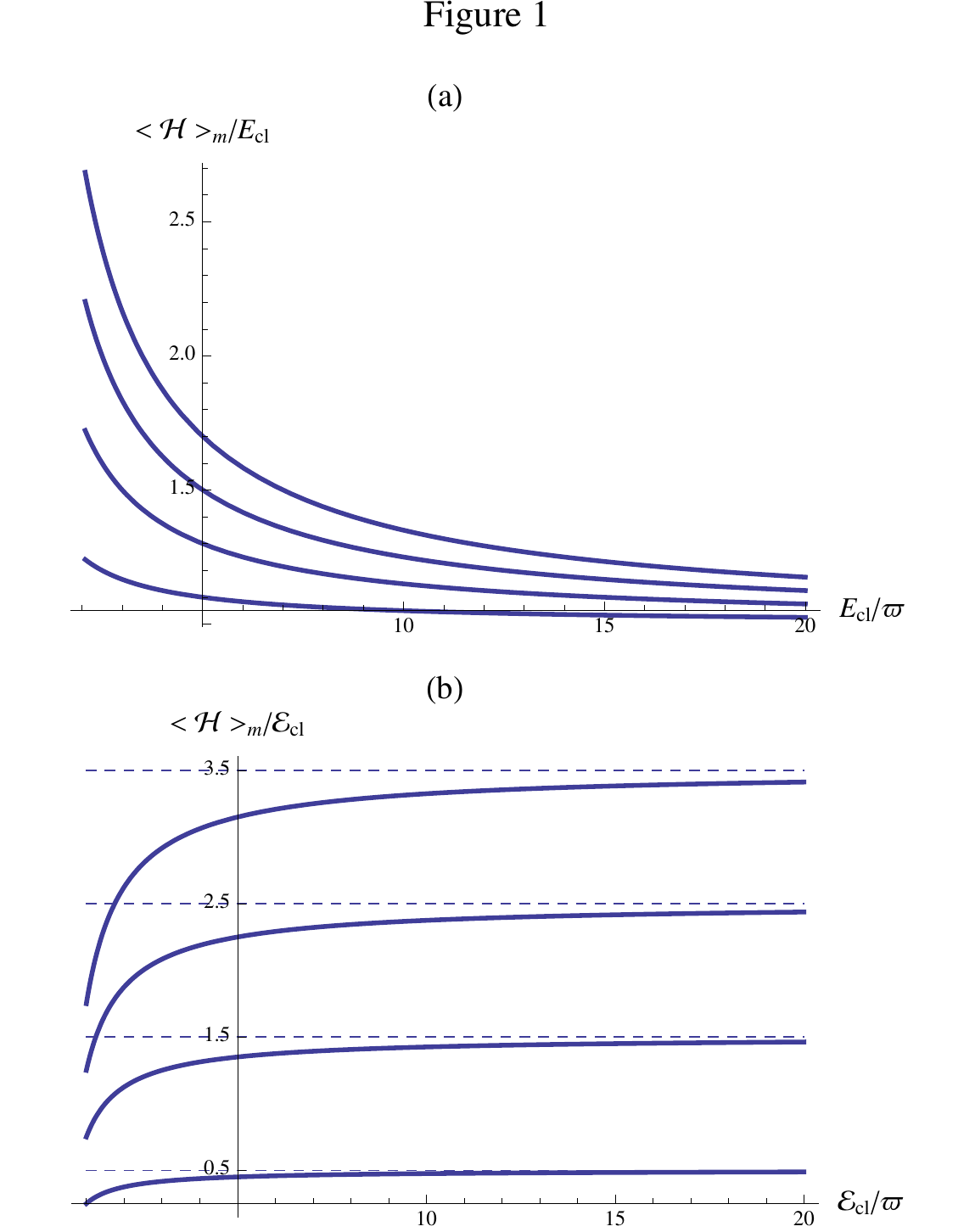}
\caption{\textbf{Asymptotic limit of the mean quantum energy $\langle\:\mathcal{H}\:\rangle_m$}. Plot of the ratio between $\langle\:\mathcal{H}\:\rangle_m$ and the classical energies $E_{cl}$ and $\mathcal{E}_{cl}$ in units of $\widetilde{\omega}$ ($\hbar=1$). \textbf{(a)}: $\rho(t)$ bounded; \textbf{(b)}: $\rho(t)$ unbounded.}
\label{default}
\end{center}
\end{figure}

\section{Choosing $\rho(t)$ unbounded and $\alpha(t)=0$.}
\label{alpha=0}

If one chooses $\alpha(t)=0$, equation~\eqref{Psitilde} reads (recall eq.n~\eqref{Psitilde-m}):

\begin{equation}
\label{Psitilde4}
\widetilde{\Psi}_m(q,\:\xi(t))=\frac{\mathrm{e}^{iB(t)}}{\sqrt{\rho}}\phi_m\left(\frac{q}{\rho}\right)\times\mathrm{exp}\left[i\frac{\dot{\rho}}{2\rho}q^2\right]\:,
\end{equation}
\\
In this case, PR and damping does enter the auxiliary function $\rho$ alone. However in analogy with the case studied in Section \ref{rhobound}, the ignition of PR on $\rho(t)$ is forbidden too, by choosing the initial state $\widetilde{\Psi}(q,\:0)$ as the \emph{exact} m-eigenstate $\phi_m\left(q\sqrt{\widetilde{\omega}(0)}\right)$ of the harmonic oscillator. Indeed, this would yield a bounded or vanishing $\rho(t)$, as shown in ref.~\cite{ArXive} and in Section \ref{LRI}. Hence, even in this case an arbitrary small perturbation, yielding $\rho(0)\ne1/\sqrt{\widetilde{\omega}(0)}$, or $\rho^{(1)}(0)\ne0$, must remove the initial quantum state from equilibrium. If so, from eq.ns \eqref{phi-n} and \eqref{Pnmgeneral}, one gets:

\begin{equation}
\label{PnmC}
P_m(n)=\frac{\sqrt{\omega(t)}}{\rho(t)}\left|\int\mathrm{d}q\:\phi_m\left(\frac{q}{\rho(t)}\right)\phi_n\left(q\sqrt{\omega(t)}\right)\mathrm{exp}\left[i\frac{\dot{\rho}}{2\rho}q^2\right]\right|^2\:
\end{equation}
\\
For spatial parity reasons, it is obvious that formula~\eqref{PnmC} vanishes, unless $m$ and $n$ are both even or odd. In the two cases, Appendix B shows that $P_m(n)$ can be expressed as:

\begin{subequations}
\label{Peven/odd}
\begin{align}
P_{2m}(2n,\:\Gamma)&=\frac{(2m)!(2n)!}{\sqrt{\Gamma}\:2^{2(m+n)}}\left(1-\frac{1}{\Gamma}\right)^{n+m}\times\nonumber\\
\nonumber\\
&\times\left[\sum_{j=0}^\mu(1-\Gamma)^{-j}\frac{2^{2j}}{(2j)!(n-j)!(m-j)!}\right]^2\label{Peven}
\end{align}
\\
\begin{align}
P_{2m+1}(2n+1,\:\Gamma)&=\frac{(2m+1)!(2n+1)!}{\Gamma^{3/2}\:2^{2(m+n)+2}}\left(1-\frac{1}{\Gamma}\right)^{n+m}\times\nonumber\\
\nonumber\\
&\times\left[\sum_{j=0}^\mu(1-\Gamma)^{-j}\frac{2^{j+1}}{(2j+1)!(n-j)!(m-j)!}\right]^2\:\label{Podd}
\end{align}
\end{subequations}
\\
($\mu=\mathrm{min}\{n,\:m\}$), in terms of the single real quantity\footnote{In passing from the first to second line in eq.n~\eqref{Gamma}, use has been made of eq.ns~\eqref{T(t)} and \eqref{Ecl}.}:

\begin{align}
\Gamma&=\frac{1}{2\omega}\frac{\left(\dot{\rho}^2+\omega^2\rho^2\right)}{2}+\frac{1}{2}+\frac{1}{4\omega\rho^2}=\nonumber\\
\nonumber\\
&=\frac{1}{2\widetilde{\omega}\beta}\underbrace{\frac{\left[
(\rho^{(1)})^2+\widetilde{\omega}^2\rho^2\right]}{2}}_{\mathcal{E}_{cl}(\rho,\:\rho^{(1)},\:t)}+\frac{1}{2}+\frac{1}{4\omega\rho^2}\:.\label{Gamma}
\end{align}
\\
If, as discussed in Section \ref{LRI}, $\rho$ oscillates with diverging amplitudes, due to PR, the classical energy $\mathcal{E}_{cl}(\rho,\:\rho^{(1)},\:t)$ diverges accordingly, and:

\begin{equation}
\label{lim}
\mathrm{lim}_{t\rightarrow\infty}\frac{\mathcal{E}_{cl}}{2\beta\widetilde{\omega}\:\Gamma}=1\:.
\end{equation}
\\
In short, if PR makes the classical energy $\mathcal{E}_{cl}(\rho,\:\rho^{(1)},\:t)$ diverge, the probabilities $P_{2m}(2n,\:\Gamma)$, $P_{2m+1}(2n+1,\:\Gamma)$ and the mean value of the quantum energy

\begin{align}
&\langle\:\mathcal{H}\:\rangle_m=\sum_{n=0}^\infty P_m(n,\:\Gamma) \beta\widetilde{\omega}[n+1/2]=\nonumber\\
\nonumber\\
&=
\begin{cases}
\sum_{j=0}^\infty P_m(2j,\:\Gamma) \beta\widetilde{\omega}[2j+1/2]&\quad\quad (m \text{ even})\\
\label{Em/even-odd}\\
\sum_{j=0}^\infty P_m(2j+1,\:\Gamma) \beta\widetilde{\omega}[2j+3/2]&\quad\quad (m \text{ odd})\:
\end{cases}
\end{align}
\\ 
depend asymptotically just on $\mathcal{E}_{cl}/(2\beta\widetilde{\omega})$, (recall eq.n~\eqref{lim}). Analytic and numerical results (Appendix B), obtained from eq.ns~\eqref{Em/even-odd} and \eqref{Peven/odd}, are shown in Fig (4b). It is seen that the asymptotic energy now is described by the formula:

\begin{subequations}
\begin{equation}
\label{<E>2}
\langle\:\mathcal{H}\:\rangle_m\rightarrow\left[m+\frac{1}{2}\right]\mathcal{E}_{cl}(\rho,\:\rho^{(1)},\:t)\quad(m=0,\:1,\:\cdots)\:,
\end{equation}
\\
while the energy's standard deviation reads, asymptotically:

\begin{equation}
\label{<sigma>2}
\langle\:\sigma_{\mathcal{H}}\:\rangle_m\rightarrow \mathcal{E}_{cl}(\rho,\:\rho^{(1)},\:t)\sqrt{[(m+1)^2-m]/2}\:.
\end{equation}
\end{subequations}
\\
\begin{figure}[htbp]
\begin{center}
\includegraphics[width=4in]{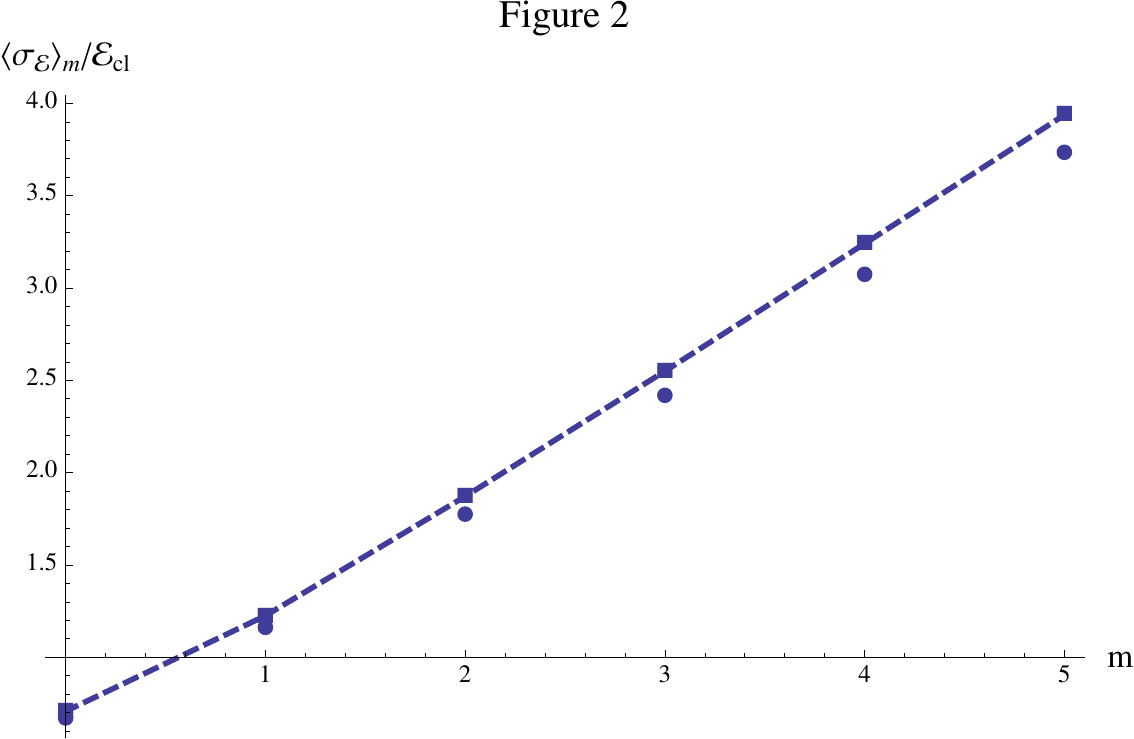}
\caption{\textbf{Ratio between energy's standard deviation and $\mathcal{E}_{cl}$ in the case $\rho(t)$ unbounded}. Circles correspond to $\mathcal{E}_{cl}/\widetilde{\omega}(t)=10$. Squares correspond to $\mathcal{E}_{cl}/\widetilde{\omega}(t)=100$. Dashed line corresponds to the asymptotic ($\mathcal{E}_{cl}\rightarrow\infty$) expression $\sqrt{[(m+1)^2-m]/2}$}.
\label{default}
\end{center}
\end{figure} 

All this shows that in the present case not only the memory of the initial state $\phi_m$ is preserved, but the classical energy $\mathcal{E}_{cl}$ plays the role of an asymptotic \textquoteleft superquantum\textquoteright$\:$of energy, generated by each initial quantum $\widetilde{\omega}(0)$. In addition, the relative quantum fluctuations $\langle\:\sigma_{\mathcal{H}}\:\rangle_m/\langle\:\mathcal{H}\:\rangle_m=\sqrt{2[(m+1)^2-m]}/(2m+1)$ converge to a finite asymptotic value, and preserve, in turn, an exact memory of the initial state (Fig. 2).

\section{Discusssion and Conclusions}
\label{concl}

In the PR theory, both classical and quantum, the LR invariant $I(p,\:q,\:\rho(t))$ (eq.n~\eqref{I}) is an important issue, as discussed in Section \ref{LRI}. Such invariant, however, requires an auxiliary function of time $\rho(t)$, satisfying the non linear equation eq.n~\eqref{eq.rho}. In ref. \cite{ArXive} the case $I=0$, is studied in detail, showing that, for an appropriate choice of the initial conditions, $\rho(t)$ is bounded anyway, and vanishes asymptotically in the presence of dissipation, despite $\omega_0\tau$ falls in a PR-band. If, in contrast, $I>0$ and $q(t)$ oscillates with diverging amplitudes (due to PR), then $\rho(t)$ oscillates with diverging amplitudes too, scaling with those of $|q(t)|$. Those different behaviors of the classical auxiliary function $\rho$ turn out to have a relevant influence on the evolution of the quantum state at long times, as shown in Sections \ref{Q}, \ref{rhobound} and \ref{alpha=0}. 

The exponential increase of the classical energy, due to PR, leads one to favor a coherent states representation of the quantum dynamics, as the most appropriate one, in view of a semi-classical asymptotic evolution \cite{PRA1992, PRA1997, MR, JB}. In turn, this choice leads to the second quantization representation as a natural consequence. The present approach, instead, adopts the Schr\"{o}dinger representation, from which it is easy to see that the auxiliary function $\rho(t)$ determines the time-dependent uncertainty $\Delta q$ of the position operator (Section \ref{Q}). According to the LR transformations, another auxiliary function $\alpha(t)$, satisfying the classical equation of motion \eqref{alpha}, yields the motion of the wave packet's centre. In specifying the details of the initial quantum state, we choose to start with states \emph{as close as possible} to the $m$-eigenstate of the initial oscillator. However, it is impossible to start up PR, unless the $m$-eigenstate is slightly perturbed. 

In the case $\rho(t)$ bounded (Section~\eqref{rhobound}), the perturbation applies to the position or velocity of the wave-packet's centre. The mean quantum energy tends to the classical diverging value $E_{cl}$, independent of $m$  (Fig. 1a), while the quantum fluctuation reads $\sqrt{\beta E_{cl}(2m+1)}$, and the relative quantum fluctuations of the energy tend to vanish. The quantum effect simply results in replacing a point in co-ordinate space, with a \textquoteleft spot\textquoteright$\:$of width $\Delta q\propto \rho$, possibly vanishing because of dissipation. No surprise, thereby, if in this case of limited or \emph{squeezed} co-ordinate uncertainty, the oscillator exhibits a \emph{fading} memory of the initial $m$-eigenstate. 

What looks more surprising is the case studied in Section \ref{alpha=0}, where the centre of the wave packet is fixed at $\alpha=0$. The auxiliary function $\rho(t)\propto\Delta q(t)$, once slightly removed from the initial equilibrium position, starts oscillating with diverging amplitudes, due to PR. In this case, the increase of the mean energy is a pure \emph{quantum} effect, driven by the time-changing uncertainty $\Delta q(t)$ of the co-ordinate. Now, not only the memory of the initial \emph{quantum} state is preserved, but the very structure of the energy eigenvalue is transferred to the asymptotic limit of the mean energy $\langle\:\mathcal{H}\:\rangle_m=\mathcal{E}_{cl}(m+1/2)$ (Fig. 4b), since each of the $m$ initial quanta produces a \textquoteleft superquantum\textquoteright$\:\mathcal{E}_{cl}$ of diverging energy. The quantum nature of the PR effect, in this case, is emphasized by the fact that the energy's standard deviation $\langle\:\sigma_{\mathcal{H}}\rangle_m=\mathcal{E}_{cl}\sqrt{[(m+1)^2-m]/2}$ diverges like the mean energy itself and depends explicitly on $m$ (Fig. 2). Hence, in the case $\rho(t)$ unbounded, in which the co-ordinate uncertainty oscillates between \emph{squeezing} ($\rho\rightarrow0$) and \emph{swelling} ($\rho\rightarrow\infty$), the memory of the initial eigenstate is \emph{persistent}, on arbitrary long time scales.

The main physical point raised by the present work is the sensitivity of quantum PR to the \emph{arbitrary small} perturbations of the initial state, that lead to quite different asymptotic behaviors, when they apply to the mean value, or to the quantum spreading of the oscillator's co-ordinate. The methods to achieve such perturbations \emph{separately} is an important technical aspect, strongly dependent on the specific applications, among those outlined in Section \ref{intro} and many possible others. What can be said, in general, is that the present study is relevant for low temperature processes, in which the oscillating system occupies the lowest energy eigenstates, and a fine control of the perturbation is made easier, by the reduced thermal fluctuations.\newline
\\           
The present work is dedicated to the memory of Harold Ralph Lewis (1931, 2002). 
  
\begin{appendices}
\numberwithin{equation}{section}

\section{Appendix A}
\label{Appendix A}

Let:

\begin{equation*}
A_m(n)=\sqrt{\langle\:\omega\:\rangle}\int\phi_n\left(q\sqrt{\langle\:\omega\:\rangle}\right)\phi_m\left((q-\alpha)\sqrt{\langle\:\omega\:\rangle}\right)\mathrm{e}^{i(\dot{\alpha}q)}\mathrm{d}q\:.
\end{equation*}
\\
On setting $x=q\sqrt{\langle\:\omega\:\rangle}$, the definitions~\eqref{Psitilde-m} and \eqref{phi-n} yield:

\begin{subequations}
\label{Pnm/c(n)/lambdaB}
\begin{align}
\label{AnmB1}
A_m(n)&=c(n)c(m)\mathrm{e}^{-\alpha^2\langle\:\omega\:\rangle/2}\times\nonumber\\
\nonumber\\
&\times\int\mathrm{exp}\left[-x^2+x\left(\alpha\sqrt{\langle\:\omega\:\rangle}+\mathrm{i}\frac{\dot{\alpha}}{\sqrt{\langle\:\omega\:\rangle}}\right)\right]H_n(x)H_m\left(x-\alpha\sqrt{\langle\:\omega\:\rangle}\right)\mathrm{d}x\:.
\end{align}
\\
where:

\begin{equation}
\label{c(n)}
c(n)=\frac{1}{(\sqrt{\pi}2^nn!)^{1/2}} 
\end{equation}
\\
Let $y=x+\lambda$, with:

\begin{equation}
\label{lambdaB}
\lambda=\frac{1}{2}\left(\alpha\sqrt{\langle\:\omega\:\rangle}+\mathrm{i}\frac{\dot{\alpha}}{\sqrt{\langle\:\omega\:\rangle}}\right)\:.
\end{equation}
\end{subequations}
\\
Thanks to the standard expression for Hermite polynomials:

\begin{equation*}
H_j(y+\lambda)=\sum_{k=0}^j\binom{n}{k}(2\lambda)^{j-k}H_k(y)\:,
\end{equation*}
\\
equation \eqref{AnmB1} becomes:

\begin{align}
A_n(m)&=c(n)c(m)\:\mathrm{e}^{-(\lambda^2+\mathrm{i}\alpha\dot{\alpha})/4}\int H_n\left(y+\lambda\right)H_m\left(y-\lambda^*\right)\mathrm{e}^{-y^2}\mathrm{d}y=\nonumber\\
\nonumber\\
&=c(n)c(m)\:\mathrm{e}^{-(\lambda^2+\mathrm{i}\alpha\dot{\alpha})/4}\times\nonumber\\
\nonumber\\
&\times\sum_{k_1=0}^n\sum_{k_2=0}^m\frac{(-1)^{m-k_2}}{c(k_1)c(k_2)}\binom{n}{k_1}\binom{m}{k_2}\lambda^{n-k_1}(\lambda^*)^{m-k_2}\:\delta_{k_1,k_2}=\nonumber\\
\nonumber\\
&=c(n)c(m)\lambda^n(\lambda^*)^m\:\mathrm{e}^{-(\left|\lambda\right|^2+\mathrm{i}\alpha\dot{\alpha})/4}\sum_{k=0}^\mu\frac{(-1)^{m-k}}{c^2(k)}\binom{n}{k}\binom{m}{k}\left|\lambda\right|^{-2k}\:,\label{AnmB2}
\end{align}
\\

Recalling the definition \eqref{Ecl}, equation \eqref{lambdaB} yields

\begin{equation*}
 |\lambda|^2=\frac{2E_{cl}(\alpha,\:\alpha^{(1)},\:t)}{\beta\langle\:\widetilde{\omega}\:\rangle}=2\Gamma\:,
 \end{equation*}
 \\
 whence, from eq.n \eqref{AnmB2}, one has:
 
 \begin{equation}
 \label{PnmB}
 P_n(m,\:\Gamma)=\left|A_n(m,\:\Gamma)\right|^2=n!m!\:\Gamma^{n+m}\mathrm{e}^{-\Gamma}\left[\sum_{k=0}^\mu\frac{(-1)^k\Gamma^{-k}}{k!(n-k)!(m-k)!}\right]^2\:.
\end{equation}
\\
From the definition of Laguerre polynomial:

\begin{equation*}
L_m^j(x)=\sum_{k=0}^{m}\frac{x^k}{k!}\binom{m+j}{m-k}\:,
\end{equation*}
\\
the r.h.s. of equation \eqref{PnmB} can be easily transformed into \eqref{Pnm2}.

The calculation of the energy's standard deviation $\langle\:\sigma_{\mathcal{H}}\:\rangle_m$, has been performed with Mathematica, by applying eq.n~\eqref{PnmB} to the first line in eq.n~\eqref{<sigma>1}. Tthe formula resulting in the second line seems to be \emph{exact}, for each $\Gamma$ and has been checked for $m=0,\:1,\:2,\:3,\:4,\:5$. This suggests that a \emph{formal} development of the expression in square bracket of eq.n~\eqref{<sigma>1} should exists, resulting in the simple formula $(2m+1)\Gamma$. However, we have not found any such simplification in the current literature on special functions.       

\section{Appendix B}
\label{Appendix B}

From the definitions~\eqref{phi-n} and \eqref{Psitilde-m}, one has:

\begin{subequations}
\label{Anm/lambdaC}
\begin{align}
A_m(n)&:=\frac{\omega^{1/4}(t)}{\sqrt{\rho(t)}}\int\mathrm{d}q\:\phi_m\left(\frac{q}{\rho(t)}\right)\phi_n\left(q\sqrt{\omega(t)}\right)\mathrm{exp}\left[i\frac{\dot{\rho}}{2\rho}q^2\right]=\nonumber\\
\nonumber\\
&=\frac{\omega^{1/4}(t)c(n)c(m)}{\sqrt{\rho(t)}}\int\mathrm{d}q\:\mathrm{e}^{-q^2\lambda/2}H_n(q\sqrt{\omega})H_m(q/\rho)=\nonumber\\
\nonumber\\
&=\frac{\omega^{1/4}(t)c(n)c(m)}{\sqrt{\rho\lambda}}\int\mathrm{d}x\:\mathrm{e}^{-x^2}H_n\left(x\sqrt{\frac{\omega}{\lambda}}\right)H_m\left(\frac{x}{\rho\sqrt{\lambda}}\right)\:,
\label{AnmC1}
\end{align}
\\
with $x=q/\sqrt{\lambda}$, and:

\begin{equation}
\label{lambdaC}
\lambda=\frac{1}{2}\left[\rho^{-2}+\omega-\mathrm{i}\frac{\dot{\rho}}{\rho}\right]\:.
\end{equation}
\end{subequations}
\\

By means of the standard formula for Hermite polynomials:

\begin{equation*}
\label{H(eta x)}
H_j(\eta\: x)=\sum_{k=0}^{[j/2]}\eta^{j-2k}\left(\eta^2-1\right)^j\binom{j}{2k}\frac{(2k)!}{k!}H_{j-2k}(x)
\end{equation*}
\\
($[j/2]=$ integer part of $j/2$), equation \eqref{AnmC1} yields: 

\begin{align}
A_m(n)&=\frac{\omega^{1/4}\:c(m)c(n)}{\sqrt{\rho\lambda}}\sum_{j=0}^{[m/2]}\sum_{k=0}^{[n/2]}\frac{\delta_{m-2j,\:n-2k}}{c^2(m-2j)}\left[\frac{\sqrt{\omega}}{\rho\lambda}\right]^{m-2j}\times\nonumber\\
\nonumber\\
&\times\left[\frac{1}{\rho^2\lambda}-1\right]^j\left[\frac{\omega}{\lambda}-1\right]^k\binom{m}{2j}\binom{n}{2k}\frac{(2j)!(2k)!}{j!\:k!}\:.\label{AnmC2}
\end{align}
\\
The contraction on Kronecker delta in eq.n~\eqref{AnmC2} forces $m$ and $n$ to be both even or odd.  A simple rearrangement of \eqref{AnmC2} yields:

\begin{subequations}
\label{Anm/even-odd}
\begin{align}
A_{2m}(2n)&=\frac{\omega^{1/4}\:c(2m)c(2n)}{\sqrt{\rho\lambda}}\left(\frac{\omega}{\lambda}-1\right)^n\left(\frac{1}{\rho^2\lambda}-1\right)^m\times\nonumber\\
\nonumber\\
&\times\sum_{j=0}^\mu\binom{2n}{2(n-j)}\binom{2m}{2(m-j)}\frac{(2(n-j))!(2(m-j))!}{c^2(2j)(n-j)!(m-j)!}\times\nonumber\\
\nonumber\\
&\times\left[\left(1-\frac{\lambda}{\omega}\right)\left(1-\rho^2\lambda\right)\right]^{-j}\:.\label{Anm/even}
\end{align}
\\
\begin{align}
A_{2m+1}(2n+1)&=\frac{\omega^{3/4}\:c(2m+1)c(2n+1)}{(\rho\lambda)^{3/2}}\left(\frac{\omega}{\lambda}-1\right)^n\left(\frac{1}{\rho^2\lambda}-1\right)^m\times\nonumber\\
\nonumber\\
&\times\sum_{j=0}^\mu\binom{2n+1}{2(n-j)}\binom{2m+1}{2(m-j)}\frac{(2(n-j))!(2(m-j))!}{c^2(2j+1)(n-j)!(m-j)!}\times\nonumber\\
\nonumber\\
&\times\left[\left(1-\frac{\lambda}{\omega}\right)\left(1-\rho^2\lambda\right)\right]^{-j}\:.\label{Anm/odd}
\end{align}
\end{subequations}
\\
From eq.n \eqref{lambdaC}, it can be shown that:

\begin{subequations}
\label{1,2,3}
\begin{equation}
\left(1-\frac{\lambda}{\omega}\right)\left(1-\rho^2\lambda\right)=1-\frac{\rho^2|\lambda|^2}{\omega}\label{1}
\end{equation}
\\
\begin{equation}
\left|\frac{\omega}{\lambda}-1\right|^2=\left|1-\rho^2\lambda\right|^2=1-\frac{\omega}{\rho^2|\lambda|^2}\label{2}
\end{equation}
\\
\begin{align}
\Gamma:=\frac{\rho^2|\lambda|^2}{\omega}&=\frac{1}{2\omega}\frac{\left(\dot{\rho}^2+\omega^2\rho^2\right)}{2}+\frac{1}{2}+\frac{1}{4\omega\rho^2}=\nonumber\\
\nonumber\\
&=\frac{E_{cl}(\rho,\:\rho^{(1)},\:t)}{2\beta\widetilde{\omega}}+\frac{1}{2}+\frac{1}{4\omega\rho^2}\:.\label{3}
\end{align}
\end{subequations}
\\
Taking the square modulus of eq.ns \eqref{Anm/even-odd}, and making use of eq.ns~\eqref{1,2,3}, it is not difficult to obtain eq.ns~\eqref{Peven/odd}.\\

An analytic calculation of $\langle\:\mathcal{H}(t)\:\rangle_m$ can be performed for the lowest values of $m$, according to eq.ns~\eqref{Peven/odd}, by recalling the identity:

\begin{equation*}
\frac{(2n)!(2\sigma)^{2n}}{2^{2n}j!}=\frac{1}{\sqrt{2\pi}\sigma}\int_{-\infty}^\infty x^{2n}\mathrm{e}^{-\frac{x^2}{2\sigma^2}}\mathrm{d}x\::
\end{equation*} 
\\
\begin{subequations}
\label{P0,1}
\begin{align}
P_0^{2n}&=\frac{1}{\Gamma^{1/2}}\frac{(2n)!(2\sigma^2)^n}{2^{2n}(n!)^2}=\frac{1}{\Gamma^{1/2}\:n!\sqrt{2\pi}\sigma}\int_{-\infty}^\infty x^{2n}\mathrm{e}^{-\frac{x^2}{2\sigma^2}}\mathrm{d}x\label{P2n,0}\\
\nonumber\\
P_1^{2n+1}&=\frac{(2\sigma^2)^n(2n+1)!}{\Gamma^{3/2}2^{2n+1}(n!)^2}=\frac{
(2n+1)}{\Gamma^{3/2}\:n!\sqrt{2\pi}\sigma}\int_{-\infty}^\infty x^{2n}\mathrm{e}^{-\frac{x^2}{2\sigma^2}}\mathrm{d}x\:,\label{P0,2n+1}
\end{align}
\end{subequations}
\\
where:

\begin{equation}
\label{2sigma2}
2\sigma^2:=\left|\frac{\omega}{\lambda}-1\right|^2=1-\frac{\omega}{(\rho|\lambda|)^2}=1-\frac{1}{\Gamma}\:.
\end{equation}
\\
With some elementary manipulations, the sums in eq.ns~\eqref{Peven/odd} can be calculated analytically in terms of integrals of the kind:

\begin{equation*}
\int_{-\infty}^\infty x^{2j}\mathrm{exp}\left[-\frac{x^2}{2\sigma^2}(1-2\sigma^2)\right]\mathrm{d}x\:,
\end{equation*}
\\
which yield, finally:

\begin{align}
\langle\:\mathcal{H}(t)\:\rangle_0&=\frac{E_{cl}(\rho,\:\rho^{(1)},\:t)}{2}+\frac{\beta\widetilde{\omega}}{4}\nonumber\\
\nonumber\\
\langle\:\mathcal{H}(t)\:\rangle_1&=\frac{3}{2}E_{cl}(\rho,\:\rho^{(1)},\:t)+\frac{5}{2}\beta\widetilde{\omega}+\cdots\nonumber\:,
\end{align}
\\
where $\cdots$ means term vanishing more rapidly than $\beta(t)$.

As for the energy's standard deviation $\langle\:\sigma_{\mathcal{H}}\:\rangle_m$, the application of eq.ns~\eqref{Peven/odd} to eq.n~\eqref{<sigma>1} leads to the numerical plot in Figure 5, showing the asymptotic limit $\langle\:\sigma_{\mathcal{H}}\:\rangle_m=E_{cl}\sqrt{[(m+1)^2-m]}$, for $E_{cl}(t)\rightarrow2\Gamma\rightarrow\infty$.

\end{appendices}

\end{document}